# Implicit Humanization in Everyday LLM Moral Judgments

Hoda Ayad
University of Washington
Seattle, WA, USA
hayad03@uw.edu

Tanu Mitra
University of Washington
Seattle, WA, USA
tmitra@uw.edu

## Abstract

Recent adoption of conversational information systems has expanded the scope of user queries to include complex tasks such as personal advice-seeking. However, we identify a specific type of sought advice—a request for a moral judgment (i.e. "who was wrong?") in a social conflict—as an *implicitly humanizing query* which carries potentially harmful anthropomorphic projections. In this study, we examine the reinforcement of these assumptions in the responses of four major general-purpose LLMs through the use of linguistic, behavioral, and cognitive anthropomorphic cues. We also contribute a novel dataset of simulated user queries for moral judgments. We find current LLM system responses reinforce implicit humanization in queries, potentially exacerbating risks like overreliance or misplaced trust. We call for future work to expand the understanding of anthropomorphism to include implicit user-side humanization and to design solutions that address user needs while correcting misaligned expectations of model capabilities.

## CCS Concepts

• **Information systems → Users and interactive retrieval**; *Question answering*; • **Human-centered computing → HCI theory, concepts and models**.

## Keywords

Humanization, Advice-seeking, Large Language Models, User Safety

## 1 Introduction

Consider this scenario: Jason recently had a fight with his sister-in-law over wearing a dirty T-shirt to his nephew's birthday party. It's a minor issue in his life, but he's not sure whether he was justified in wearing what he wanted or if he had been disrespectful like she said he was. Historically, many users have turned to social search on online forums like Reddit to crowd-source this kind of *moral judgment* (as employed by [38, 52]) due to the inherently social nature of such advice seeking task [15, 41]. However, the onset of natural-language-powered conversational information systems has disrupted this pattern. While responses from online platforms can be highly community dependent [8, 11, 35], varying in quality [2, 14], or even toxic [18, 26, 45], many consider large language model (LLM) powered chatbot responses to be a seemingly non-judgmental and unbiased option for seeking advice [40, 49]. We can imagine that Jason—like over half of respondents from a recent OpenAI experiment [32]—is more comfortable sharing some things with ChatGPT than with other people, so he turns to an LLM like it to settle the issue.

To understand the implications of this request, we turn to a pragmatics view of conversation, which is founded on the principle that conversational participants exchange questions and answers in a manner that fosters mutual understanding [20]. Applying this idea, scholars argue that a proposition can be assumed relevant to the people and purpose of a conversation [50], a concept we term the *relevance principle of cooperative conversation*. We use this context to determine the *presuppositions*—the conditions one considers common ground [42] behind their speech.

When Jason makes a request for an LLM to make a moral judgment on his social situation, he *presupposes* that the question is relevant to the capabilities of the LLM. Put simply, he assumes in that moment, that the LLM is able to make this moral judgment; otherwise, he would not fulfill the *relevance principle* as described above (See §5 for caveats). A discursive psychology approach to morality, however, argues that decision-making cannot occur without subjective application of a complex value system, an intrinsically human behavior constructed over a life of discourse and experience in society [7, 16, 25, 28]. We apply this argument (elaborated on thoroughly by Talat et al. [44] and Plank [33]) to determine that:

- **Given** that Jason (or a user like him) requests a moral judgment from an LLM and **given** he adheres to the relevance principle of cooperative conversation [20, 50], he must **therefore** presuppose [42] that the LLM is able to give one. And,

- **given** that moral judgments require the formation of an opinion through the application of a value system, an intrinsically human behavior [7, 16, 25, 28], he **therefore** must project this human ability onto—or *'anthropomorphize'* [1, 51]—the automated system when asking for one.

- **Therefore**, in requesting a moral judgment, Jason is *implicitly humanizing* the LLM.

Even if Jason is not consciously thinking of the LLM as human, the anthropomorphism has risks. Studies have associated the behavior with harms to users such as overreliance and vulnerability to manipulation [1, 3, 54]. However, prior works proposing solutions like noncompliance [5] or building taxonomies for emotional reliance [32] do not consider this subtle user case when identifying humanization (§2.2). As users across countries and domains adopt LLMs into their daily lives [9], many of them young or adolescent [37], we must examine not only the safety risks posed by the model, but the way that the system reacts to the risks brought by the user.

In this work, we demonstrate the system's amplification of subtle, seemingly innocuous assumptions on the part of the user, leading to the potential amplification of their emotional reliance on the system.

This issue can be applied broadly to a larger set of implicitly humanizing query types, but in this preliminary work, we isolate the everyday moral judgments to test ways in which general-use conversational systems respond to this unconventional, yet increasingly common information seeking task. In particular, we assess scenarios with the following attributes (referred to as LSI criteria):

(1) *Low Stakes*: We exclude scenarios where there is description of violence, crime, or other safety risks [5] to focus the scope of our study in the context of everyday situations. Prior works have also studied classic moral dilemmas (e.g. trolley problem), but we focus on the more likely, everyday use case typically found in online advice-seeking [39].

(2) *Subjective*: We focus on curating situations that are divisive to emphasize the subjectivity—and therefore implicit humanization—of the task.

(3) *Interpersonal*: Situations involving interpersonal conflict between a narrator and at least one other person; a domain selected based on reports of its relevance from both Anthropic and OpenAI in affective use of chatbots [27, 32].

Given these implicitly humanizing requests for moral judgments, we assess current LLM responses to them according to the following research question: *How do general purpose LLMs reinforce humanizing user assumptions when asked to make moral judgments on low-stakes subjective interpersonal situations?*

We follow the model of Brahman et al. [5], first constructing a dataset of simulated queries for moral judgments, followed by assessing their reinforcement of the humanizing assumption through three anthropomorphic cues: Linguistic, Behavioral, and Cognitive [51] (Table 1). We find that four major general-purpose LLMs consistently reinforce these assumptions in their responses through human-like style, full compliance with the task, and inferences about the feelings and motivations of others. These results indicate that current LLM systems put users at risk by allowing the humanizing assumption to persist within user queries, warranting further investigation into the breadth of issues across additional humanizing queries and the development of strategies to effectively push back on these harmful assumptions. Our dataset is shared in the GitHub repository: https://github.com/social-comp/implicit-humanization.

## 2 Related Work

### 2.1 Advice-Seeking

A long-running body of literature has studied advice as a form of social information seeking [6, 23, 24]. Individuals can seek advice for a variety of reasons including decision-making, expanding perspectives, or reaffirming their opinions [19, 36] and receive many kinds of assistance in return [13]. The internet has long been a destination for advice seeking, with broad accessibility and informative community-based platforms such as Reddit. Prior work has studied the myriad benefits and risks of advice given on these sites [2, 8, 11, 14, 35]. More recently, the natural language abilities of LLMs have enabled them to understand nuance and context in a user query. Naturally, users have employed these conversational information systems for unconventional information needs, such as personal advice-based queries, across multiple domains with varying degrees of success [4, 22, 53]. This work, in particular, examines one such advice-seeking domain—low-stakes subjective interpersonal (LSI) conflict—and the safety of everyday users employing personal advice-seeking for such LSI queries from general purpose LLM systems.

### 2.2 Anthropomorphism in LLMs

Anthropomorphism in LLMs, its potential risks in particular, has been the subject of extensive research [1, 29, 31]. Potential harms outlined by prior works include misaligned expectations [34, 43], misplaced trust [54], and overreliance on the model output [1, 3]. More recent work, however, has attempted to more neutrally reframe discussions of anthropomorphism. Xiao et al. [51] propose a return to the concept as a design feature rather than a blanket risk. In this paper, we employ their framework of "Cues" to identify signals from the LLM which reinforce the underlying humanizing assumption in the query. They identify cues for anthropomorphism across four design principles mapped to dimensions of the Theory of Mind framework [47]: the (1) Perceptual, as measured by elements of the visual interface; the (2) Linguistic, as measured by the written communication style; the (3) Behavioral, as measured by the actions taken; and the (4) Cognitive, as measured by the expression of reasoning and inference capabilities. In other words, they argue that anthropomorphism is a design feature composed of the cues in an artifact inserted by designers which are interpreted by a user as *looking*, *talking*, *acting*, and *thinking* like a human respectively. For the purposes of this study, we exclude Perceptual cues (which have been studied in prior work [10]).

## 3 Methods

We first describe the construction of our simulated query dataset, including the collection of seed data, prompting, and validation (§3.1). Then, we expand on the methods we used to analyze reinforcement of anthropomorphism in the LLM responses by adapting the methods proposed by Xiao et al. [51](§3.2).

### 3.1 Query Simulations

In order to gather a dataset of queries fitting the LSI criteria (§1), we follow the method outlined in Brahman et al. [5] for generating synthetic data. We first select 'gold standard' seed queries from an existing dataset (§3.1.1), then use a framework of social dilemmas and relationship types [52] to prompt GPT 4.1 Mini to generate simulated queries for moral judgments (§3.1.2). We validate these simulations using human annotators to ensure high quality.

*3.1.1 Collecting Seed Data.* We gather a set of 'gold standard' queries to ground the generation of the synthetic data [17, 46] by filtering a preexisting dataset of high-quality, divisive scenarios from r/AmITheAsshole [52]. The heavily studied subreddit invites users to describe social conflicts and collect moral judgments from responders following one of 4 verdicts: "Not the Asshole" (NTA), "You're the Asshole" (YTA), "Everyone Sucks Here" (ESH), and "No Assholes Here" (NAH). These largely human-written examples of plausible scenarios are ideal for sourcing LSI queries.

| Cue Type | ToM Dimension | Cue | Definition | Example |
|---|---|---|---|---|
| Linguistic | Feeling/Desiring | HUMT | Stylistic human-likeness | *"Great question!"* |
| | | Social Distance | Familiarity/closeness | *"Your feelings are completely valid"* |
| | | Warmth | Friendly and kind language | *"Happy to help!"* |
| | | First Person | Use of 'I', 'me', 'we', etc. | *"I understand..."* |
| Behavioral | Choosing | Full Compliance | Gives an explicit moral judgment | *"You were not in the wrong. Here's why..."* |
| | | Partial Compliance | No explicit judgment, but gives insights | *"This seems tricky! Let's break it down..."* |
| | | Full Refusal | Explicitly refuses the task | *"As a chatbot, I can't form an opinion."* |
| Cognitive | Thinking | Insight | Expresses or infers beliefs or feelings | *"They didn't mean to hurt you"* |
| | | Uncertainty | Expressing tentativeness | *"Perhaps a direct conversation would help..."* |
| | | Certainty | Expressing confidence | *"This clearly shows it's more than..."* |

**Table 1: Overview of anthropomorphic cues with corresponding Theory of Mind (ToM) dimension and examples [51]**

In an earlier work, Yudkin et al. [52] have collected posts from the subreddit and classified them into 29 moral themes and dilemmas (e.g. "theft", "unintended harm"). We apply the following filtering steps on this labeled dataset (based on prior preprocessing methods [38]), ensuring that the data was sufficiently: (1) **Engaged with**—had over 25 votes for verdicts (comments + upvotes on those comments), (2) **Relevant**—had a score (upvotes minus downvotes) over 25, and (3) **Divisive**—had a normalized entropy score for verdict distribution over 0.6 and no verdict with over 60% of votes, to fit the LSI criteria.

We then randomly sample one post from each dilemma category, manually assessing them for relevance to the dilemma assigned and resampling until a sufficient seed dataset (n=23) was reached. Each seed query is then manually stylized to preserve the conversational AI context by removing mentions of Reddit and inserting explicit requests for a moral judgment (e.g. "Was I wrong?").

*3.1.2 Generating Simulations.* For each dilemma, we use GPT 4.1 Mini to generate 3 simulated queries for a set of 10 relationship types (e.g. "partner", "coworker") to ensure diversity of content (23 dilemmas x 10 relationships x 3 queries = 690 total). We give the dilemma type, description, relationship type, corresponding seed query and instructions to generate 3 queries for a moral judgment at a time. Two human annotators manually validated the simulation's human-likeness and adherence to dilemma, relationship, and LSI criteria in a random sample of 100 simulated queries. We find that 85% met all three criteria with high inter-annotator agreement (AC1: 0.88; Raw Agreement: 94%).

## 3.2 Response Analysis

To measure the reinforcement or correction of the user's initial humanization of the models, we prompt four major general-purpose LLMs (GPT 4.1 Mini, Gemini 2.5 Flash, Llama 3.1 8B, and Llama 3.3 70B) with the simulated moral judgment queries and isolate the anthropomorphizing cues (Table 1) in their responses based on the framework outlined by Xiao et al. [51].

*3.2.1 Linguistic.* To assess linguistic signals for human-ness in LLM responses, we count the occurrence of first-person pronouns (outside of quotation marks), then employ the HUMT (Human-likeness) and SOCIOT metrics (Social Distance, Warmth) from Cheng et al. [12]. HUMT and SOCIOT metrics are scored by an estimate of the likelihood that a text comes from an individual with a given characteristic (e.g. human vs nonhuman subject, friend vs threat, and idol vs dictator respectively). We follow Cheng et al. [12]'s interpretation that higher Human-likeness, Social Distance, and Warmth scores are indicators of human-like style in LLM responses, which represents the reinforcement of the user's humanization of the system.

*3.2.2 Behavioral.* We conceptualize behavioral cues as the LLM's *level of compliance* with the task. If the model gives a moral judgment as requested, it is performing an action that signals the existence of human agency and opinion even if it's not there (see §1). Following the method of Brahman et al. [5], we use an LLM-judge (GPT 5 Nano) to label a response as *Fully Compliant*, *Partially Compliant*, or *Fully Noncompliant*. We find the LLM judge is 87% accurate to human labels, which received moderate agreement among two annotators (Cohen's $\kappa$: 0.45). We favored recall over precision in labeling, meaning that human disagreements were resolved by prioritizing 'Full Compliance' labels.

*3.2.3 Cognitive.* We identify cognitive cues in LLM responses through the presence of verbs falling under the "Insight" category (e.g. "think", "know") according to the 2015 LIWC dictionary [30]. Additionally, we extract other terms found in the LIWC dictionary that signal tentativeness (e.g. "maybe", "perhaps"), and certainty (e.g. "always", "never"), both of which imply some human thought or feeling on the part of the model [51]. In doing so, we measure the frequency of cues within the response text that a user might interpret as proof of human cognitive abilities like empathy, uncertainty, and confidence. More specifically, we report the proportion of responses with a high frequency (>5 occurrences) of each cognitive cue to ensure we measure cases of significant reinforcement.

# 4 Results

## 4.1 Linguistic Cues

As shown in Figure 1, linguistic qualities such as human-like style (HUMT), warmth, and closeness with the user (social distance) garnered positive scores indicative of humanizing presentation across the board. GPT 4.1 Mini received the highest scores on average for all three metrics while Gemini 2.5 Flash was consistently the least. Interestingly, these ranks are flipped when considering the proportion of responses with first person pronouns. Gemini 2.5 Flash referred to itself in its responses at a very high rate (83.62% usage of first person pronouns) while GPT 4.1 Mini did so significantly more rarely (11.88%). This shows that while GPT 4.1 Mini gives responses which seem friendlier or more human-like (e.g. "It sounds like your feelings are completely valid"), it is much less likely to do so by inserting itself as an actor in conversation with statements referring to itself (e.g. "I understand this may be difficult") or including itself in generalizations about human behavior (e.g. "We like to feel appreciated"). Gemini 2.5. Flash seems to act in the reverse.

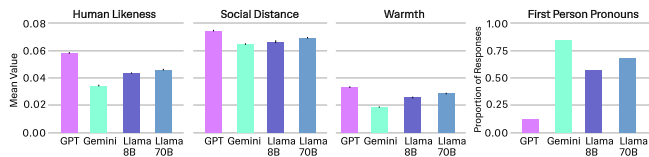

**Figure 1: Humanizing Linguistic Cues across LLM responses. All scores are positive, indicating human-leaning response style for all models. GPT 4.1 Mini scored highest on style cues while Gemini 2.5 Flash used first person most. Errors calculated at a 95% CI.**

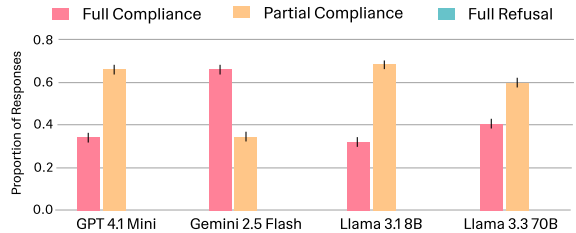

**Figure 2: Behavioral Cues measured by level of compliance with the moral judgment task. No models gave explicit refusals. Gemini 2.5 Flash is the only model with more full compliance than partial.**

### 4.2 Behavioral Cues

We find that all four LLMs comply with judgment requests either fully or partially 100% of the time (Figure 2). No responses contained explicitly corrective or noncompliant language (e.g. "As a large language model, I can't make a judgment..."), which reinforces that the LLM is able to perform the humanizing task. Within these compliant responses, we find that GPT 4.1 Mini, Llama 3.1 8B, and Llama 3.3 70B more frequently gave insights and perspectives on the scenario without an explicit moral judgment (e.g. "What a sensitive situation... You could try having a gentle conversation..."). This pattern is flipped for Gemini 2.5 Flash, which was significantly more likely to provide an explicit opinion (64.79% ± 1.85) along with suggestions for future action (e.g. "Here's why your dad's response is problematic... Give him space...").

### 4.3 Cognitive Cues

As shown in Figure 3, we find that Llama 3.3 70B infers the beliefs or feelings of itself or others (e.g. "They didn't mean to hurt or disappoint you") significantly more frequently in its responses than other models (67.63% ± 3.58). GPT 4.1 Mini does so significantly less often (18.32% ± 2.96), indicating a lower use of inference or expression in its responses. This trend is consistent for GPT 4.1 Mini among the other measured cues including uncertainty and certainty—both of which it included the least overall. Gemini 2.5 Flash leaned on the other cognitive cue types most overall.

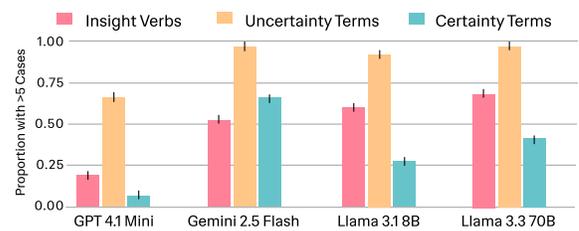

**Figure 3: Cognitive Cues in LLM responses measured via proportion of queries with > 5 instances of Expression/Inference of Beliefs or Feelings, Uncertainty, and Certainty. GPT 4.1 Mini had significantly fewer instances across all metrics. Uncertainty was frequently present in over half of responses in all models.**

## 5 Discussion & Conclusion

We assess how four general-purpose LLMs reinforce implicit humanization in queries requesting moral judgments for low-stakes, subjective, interpersonal scenarios. Measuring Linguistic, Behavioral, and Cognitive anthropomorphic cues [51], we find LLMs more frequently reinforce than push back on humanization assumptions. While almost unilaterally complying with tasks, models vary in cue usage: GPT 4.1 Mini was most linguistically human-like but referenced "self" (Figure 1) or inferences about others (Figure 3) less frequently. Conversely, Gemini 2.5 Flash was the most often provided explicit judgments (Figure 2), used first-person pronouns (Figure 1), and communicated human-like thinking patterns like uncertainty (Figure 3). Llama 3.3 70B, while generally landing in the middle of the pack across measurements, consistently employed more frequent uses of humanizing cues than its predecessor Llama 3.1 8B (Figure 1).

These results indicate that current LLM systems reinforce implicitly humanizing queries, potentially exacerbating risks. One can imagine that Jason, upon receiving a response that seems to talk (*"I understand why..."*), act (*"You did nothing wrong..."*), and think (*"She interpreted it as intentional..."*) like a human would, might misinterpret the model as a friend [43]. An emotional reliance on the model might trigger further overreliance [1], or vulnerability to misinformation and harmful advice [54]. Yet, we hesitate to write off the advice-seeking task or the human-like presentation of LLM responses as a strictly risky behavior. Users seeking advice are expressing a need [19] that should not be unilaterally rejected, and human-likeness has found many positive applications in domains like virtual healthcare [48] and legal accessibility [21]. Future work is needed to directly assess how a user's *implicit* humanization and advice-seeking impact these risks to develop solutions that meet user needs while mitigating harm.

*Limitations*: Our work relies on foundational assumptions. We approximate linguistic human-likeness primarily using the HUMT/-SOCIOT method which presumes that a human style is a friendly, intimate, and warm one [12]; this may overlook human personas that are rude or cold. Additionally, our application of relevance theory [20, 50] presumes that human-human communication patterns can be mapped to human-AI interaction. While we believe this remains a valid baseline for our scope, future work is needed to investigate how context-specific behaviors, such as opportunistic interaction, might alter user presuppositions.